# Entropic forces in Brownian motion

Nico Roos*
*Leiden Observatory, PO Box 9513, NL-2300 RA Leiden, The Netherlands*

Interest in the concept of entropic forces has risen considerably since E. Verlinde proposed in 2011 to interpret the force in Newton's second law and gravity as entropic forces. Brownian motion—the motion of a small particle (pollen) driven by random impulses from the surrounding molecules—may be the first example of a stochastic process in which such forces are expected to emerge. In this note it is shown that at least two types of entropic force can be identified in three-dimensional Brownian motion (or a random walk). This analysis yields simple derivations of known results of Brownian motion, Hooke's law, and—applying an external (non-radial) force—Curie's law and the Langevin-Debye equation.

## I. INTRODUCTION

An entropic force is an emergent phenomenon resulting from the tendency of a thermodynamic system to maximize its entropy. The system, or the set of macroscopic variables describing it, tends to evolve from one state to another state that is statistically more probable.[1] The evolving system appears to be driven by a force. A well-known example is the contraction of a polymer in a warm bath.[2] The concept of entropy (along with temperature) has also become important in gravitational physics and cosmology. Recently E. Verlinde[3] has successfully applied the concept of entropic forces to give new interpretations of Newton's second law and gravity. The observed cosmic acceleration appears to be quantitatively in agreement with an entropic force exerted by the horizon of the Universe,[4] although recent observations of the Hubble expansion seem to be better explained by the enigmatic cosmological constant.[5]

In this paper we will study the role of entropic forces in classical Brownian motion.[6] We will see that the results can be applied to derive Hooke's elasticity law and Curie's law for a magnet in a warm bath. Descriptions of Brownian motion can be found in the first chapters of most textbooks on statistical mechanics. The early studies of this process by Einstein, Smoluchovski, Langevin, and Ornstein and Uhlenbeck form the basis of the large field of stochastic processes in physics and mathematics.[7] It is surprising that it is difficult to find examples in the literature where the concept of an entropic force is applied to Brownian motion. To my knowledge, R. M. Neumann is the only one who has made an attempt to do this.[8] Starting from Boltzmann's equation he derives a radial entropic force driving the particle away from a fixed point. His derivation of the famous Einstein relation,[9]

$$\langle x^2 \rangle = 2Dt, \qquad (1)$$

relating the mean-square displacement to the product of the diffusion coefficient and time, is short and straightforward. In this paper I will draw attention to this approach and illustrate how simple arguments based on the principle of minimum information give rise to entropic forces such as Neumann's "radial force," as well as a tangential force which becomes apparent once a preferred direction is chosen or when an external (non-radial) force is applied.

Equation (1) is similar to the well-known relation for the random walk, often referred to as the drunkard's walk. Suppose that at closing time a large number of drunkards leave the pub. They all start walking randomly around the pub. Then the mean distance of the drunkards from the pub is related to the time since they left by $\langle r \rangle \propto t^{1/2}$. The drunkards seem to be pulled away from the pub by a mean macroscopic force that is proportional to $\langle r \rangle^{-1}$. Neumann calls this the radial force and explains it as the result of the tendency of the system to increase its entropy by evolving towards a state with a larger number of microstates. In a spherically symmetric (3D) system, the number of configurations $\Omega$ in a shell of thickness $dr$ is proportional to $r^2$. Using Boltzmann's expression for the entropy, $S = k_B \log \Omega$, Neumann finds for the mean radial entropic force

$$\langle F_r \rangle = T \left\langle \frac{dS}{dr} \right\rangle = \frac{2 k_B T}{\langle r \rangle}. \qquad (2)$$

In this expression, which will reappear in Section III (Eq. (20)), $k_B$ is Boltzmann's constant and $T$ is related to the random speed of the drunkard. In the case of a Brownian particle, $T$ is the temperature of the surrounding molecules. Equation (1) then follows by combining this equation with Stokes' law (see below).

In the derivation of the radial (and tangential) entropic force on a Brownian particle we follow Verlinde.[3] He considers a system consisting of a polymer embedded in a heat bath in the microcanonical ensemble, with total multiplicity $\Omega(E + F_r r, r)$, to find $F_r$, the entropic force on a polymer molecule that is stretched over a distance $r$. Here $\Omega$ is the number of microstates of the entire system (polymer plus heat bath) as a function of the total energy $E + F_r r$, where $E$ is the energy of the heat bath and $F_r r$ is the energy applied by the (radial) entropic force. In the microcanonical ensemble the total energy remains constant. The radial entropic force follows from Boltzmann's law and the extremal condition for the entropy,

$$\frac{d}{dr} S(E + F_r r, r) = 0 , \qquad (3)$$

yielding

$$F_r = T \frac{\partial S}{\partial r}, \qquad (4)$$

with $T^{-1} = \partial S / \partial E$ .

In Section II we will calculate the particle distribution from the principle of minimum information. In Section III the results are used to derive some basic properties of Brownian motion. In Section IV the close connection between the entropic force on a Brownian particle and the entropic force on a freely jointed chain and Hooke's law are discussed. In Section V it is pointed out that the introduction of a preferred direction, determined, for instance, by an external force, yields another component to the energy in the microcanonical ensemble, defining a tangential component to the entropic force. The results are applied to calculate the equilibrium of a pulled chain of freely jointed monomers, as well as the orientation of magnetic (or electric) dipoles in a thermal bath with an external magnetic (or electric) field.

## II. RANDOM WALK FROM THE CENTER OF A SPHERE

The first step in the derivation of the entropic force is to find an expression for the distribution of drunkards around the pub, or, in three dimensions, the distribution of Brownian particles that are released from a central point. Brownian motion is a limiting case of a random walk.[10] It is an example of a Wiener process, a class of well-studied stochastic processes. There are several ways to find its basic properties. One usually invokes the central limit theorem to derive the distribution. Our derivation will be based on maximizing the entropy, or the amount of missing information (Ref. 11) we have on the position of the particles undergoing stochastic motion. We will do this by dividing the available volume in small equal size cells and noting that, in the absence of any further knowledge about the distribution of particles, the probability to find a particle in a given cell is proportional to the volume of the cell. This gives a probability distribution $Q$. Next we look at an arbitrary distribution $P$ of $N$ particles over the cells and ask what the probability is that this distribution will occur given the fact that all cells have equal probability. The sought for distribution is then the most probable (or maximum entropy) distribution that is consistent with some trivial constraints from the random process.

The distribution $Q$ has radial symmetry and a maximum radius $r_{max} = v_0 t$, where $v_0$ is the speed of the particles (step length divided by the time per step) and $t$ is the time since release. For this distribution the probability $q(r,t)dr$ to find a particle in the shell between $r$ and $r + dr$ at time $t$ is proportional to the volume of the shell, or $q(r,t)dr \propto 4\pi r^2 dr$ for $r < r_{max}(t)$, where the normalization constant depends on time. Let us divide the sphere into a finite number of shells $k_{max}$ with thickness $\Delta r = r_{max}/k_{max}$, and replace $q(r,t)$ by $q_k \propto 4\pi(k/k_{max})^2$, the probability to find the particle in the $k$th shell. This is the discrete distribution we would adopt if we had no further information besides spherical symmetry and a maximum radius (or $k_{max}$). There are many other ways to distribute the particles over the $k_{max}$ shells. We are looking for the distribution $P$ given by $p_k$, the probability to find a particle in the $k$th shell, that is most likely to occur, given the fact that it is spherically symmetric and consistent with some trivial pieces of information about the random walk. Any other choice for $p_k$ is unjustified, because it would contain unwarranted information.

If we distribute $N$ particles spherically symmetric over the $k_{max}$ shells, the probability that they will be distributed according to $p_k$ is given by the multinomial distribution,

$$\text{Prob}(P) = N! \prod_{k=1}^{k_{max}} \frac{q_k^{Np_k}}{(Np_k)!}. \tag{5}$$

Using Stirling's formula for $\log N!$ we find

$$\log \text{Prob}(P) \cong \text{const} - \sum_k p_k \log\left(\frac{p_k}{q_k}\right). \tag{6}$$

We are interested in the second term, which is known as the Kullbach-Leibler divergence (Ref. 12), also called the relative entropy,

$$D(P|Q) = \sum_k p_k \log\left(\frac{p_k}{q_k}\right). \tag{7}$$

It is a (non-negative) measure for the difference between two distributions. Note that for a one-dimensional uniform distribution $-D(P|Q)$ is equal to the Shannon entropy.[13] Maximizing $\log \text{Prob}(P)$ is equivalent to minimizing $D(P|Q)$. It should yield a distribution $p_k$ that is consistent with our lack of information about the whereabouts of the particles, in accordance with the principle of minimum information.

We have some trivial pieces of information about the random walk process that can be incorporated, using Lagrange multipliers, when we maximize $\log \text{Prob}(P)$. We know that the particles have traveled a finite distance, so the normalized ($\sum_k p_k = 1$) distribution should have finite moments. Symmetry tells us that the first moment should be zero: $\sum_k \boldsymbol{k} p_k = \langle \boldsymbol{k} \rangle = 0$. The variance $\sum_k k^2 p_k = \langle k^2 \rangle$ should be finite, where $k^2 = \boldsymbol{k} \cdot \boldsymbol{k} = k_x^2 + k_y^2 + k_z^2$. Maximizing Eq. (6) using Lagrange multipliers to incorporate these constraints, we find that the distribution should have the form $p_k \propto k^2 \exp(-k^2/\sigma(t)^2)$, or, for a continuous distribution, $p(r,t)dr \propto r^2 \exp(-r^2/\sigma(t)^2)dr$, where $p(r,t)$ is the probability to find a particle between $r$ and $r + dr$ after some time $t = N\Delta t$ or after $N$ random steps of length $\Delta x$ (duration $\Delta t$). Finally we find for the normalized distribution

$$p(r, r_{max}, t)dr = \left[\text{erf}(r_{max}/\sigma) - \frac{2}{\sqrt{\pi}}\frac{r_{max}}{\sigma}e^{-r_{max}^2/\sigma^2}\right]^{-1} \frac{4}{\sqrt{\pi}}\sigma^{-3}r^2 e^{-r^2/\sigma^2} dr. \qquad (8)$$

The normalization factor inside square brackets goes to 1 for large $N$ or $t$ (see below) and the distribution approaches the well-known Gaussian form

$$p(r, t)dr = \frac{4}{\sqrt{\pi}}\sigma^{-3}r^2 e^{-r^2/\sigma^2} dr, \qquad (9)$$

or, with $\boldsymbol{r} = \vec{r}$ and $p(r,t)dr = P(\boldsymbol{r},t)d\boldsymbol{r}$,

$$P(\boldsymbol{r}, t)d\boldsymbol{r} = (\pi\,\sigma(t))^{-3/2}\, e^{-r^2/\sigma(t)^2}\, d\boldsymbol{r}, \qquad (10)$$

where $\sigma = \sigma(t)$ is readily shown to be equal to $\frac{1}{2}\sqrt{\pi}\langle r\rangle$ and the variance of the distribution is $\frac{1}{2}\sigma(t)^2 = \frac{1}{3}\langle r^2\rangle = \langle x^2\rangle$. There is a simple geometrical explanation for the factor 2/3 in $\sigma(t)^2 = \frac{2}{3}\langle r^2\rangle$ (see the end of this section. The density distribution can be interpreted as a (normalized) particle density distribution (or concentration) for a large number of particles or as a probability density distribution in the case of a single (Brownian) particle. The probability distribution (9) consists of two factors. The exponential factor is the probability distribution in Cartesian coordinates. It is proportional to the number of ways the particle can go from the origin to $\boldsymbol{r}$ and is proportional to the concentration of particles when a large number of particles is released from the origin. The other factor, which is proportional to $r^2$, represents the increase with $r$ of the volume elements in spherical coordinates. We will see below that the first factor is directly related to the osmotic force and to the elastic force in Hooke's law, whereas the second factor gives rise to Neumann's radial entropic force on an individual Brownian particle at $r$.

Now suppose that the particle takes $N$ steps of length $l$ at time intervals $\delta t$, so $N = L/l = t/\delta t$, where $L$ is the total path length. A simple scaling argument tells us that $\sigma(t)$ should scale as $l$ or $\delta t$ since the form of the distribution should stay the same when we change $l$ or $\delta t$ by some factor. The expression for the variance in terms of these variables is found from

$$\frac{1}{2}\sigma(t)^2 = \langle x^2\rangle = \sum_i^N l^2 \cos^2\theta_i = Nl^2\langle\cos^2\theta\rangle \qquad (11)$$

$$\Rightarrow Nl^2 \frac{1}{2}\int_{-1}^{1}\cos^2\theta\, d\cos\theta = \frac{1}{3}Nl^2 = \frac{1}{3}Ll$$

$$= \frac{1}{3}t(l^2/\delta t),$$

where $\theta_i$ is the angle between the direction of the $i$th step $\boldsymbol{r_i}$ and the $x$-axis. This is the well-known result from the theory of random walks, that the distance increases with $\sqrt{N}$ or $\sqrt{t}$. Note that the distribution (8) approaches Eq. (9), since $(r_{max}/\sigma(t))^2 \propto N \propto t$.

It is interesting to note that the increase of $r = |\boldsymbol{r}|$ with time is due to motions in the plane perpendicular to $\boldsymbol{r}$. This is due to the fact that the motions along the direction $\boldsymbol{r}$ do not change the mean value of $r$, whereas each step in the plane perpendicular to $\boldsymbol{r}$ leads to an increase in $r$. About 2/3 of all the possible motions of a particle starting at $\boldsymbol{r}$ cause an increase in $r$. This can be seen by ignoring steps in the radial direction and looking only at the tangential component of a single step made by a particle at $\boldsymbol{r}$. The average step length in the plane perpendicular to $\boldsymbol{r}$ is given by $l_{tang}^2 = \frac{2}{3}l^2$. The increase in $r$ as a result of this step is given by $(r + dr)^2 = r^2 + l_{tang}^2$, or $2rdr = l_{tang}^2$ for $r \gg l_{tang}$. The increase of $r$ with time is given by $r^2 = l_{tang}^2 t/\delta t = \frac{2}{3}l^2 t/\delta t$, which is equal to $\sigma(t)^2$ according to Eq. (11). So, the random steps in the radial direction account for $1/3$ to the increase in $\langle r^2 \rangle$, but they do not increase $r$.

## III. BROWNIAN MOTION, DIFFUSION, AND THE RADIAL ENTROPIC FORCE

So far we have only used geometrical arguments and the principle of minimum information. The connection with physics is suggested by the observation that Eq. (11) is a solution of the differential equation

$$\sigma(t)\frac{d\sigma(t)}{dt} = 2D, \tag{12}$$

where $D = l^2/(6\delta t)$ is a constant. In the case of Brownian diffusion or osmosis we will get an identical expression identifying $D$ with the physical diffusion constant. With Eq. (12) we see that the distribution we found (Eq. (10)) is a solution of the diffusion equation in radial coordinates. It is the Green's function for the diffusion equation, given by

$$\frac{\partial P}{\partial t} = D\left[\frac{\partial^2 P}{\partial r^2} + \frac{2}{r}\frac{\partial P}{\partial r}\right]. \tag{13}$$

This diffusion equation can also be derived from Eqs. (12) and (10) and the requirement that mass (or probability) is conserved. This requirement is expressed by the continuity equation,

$$\frac{\partial P}{\partial t} + \frac{1}{r^2}\frac{\partial}{\partial r}(r^2 P v_r) = 0, \tag{14}$$

where $v_r$ is the radial velocity. Inserting Eq. (10) and solving the differential equation, we find

$$v_r = \sigma^{-1}(t)\frac{d\sigma(t)}{dt}r = \frac{2Dr}{\sigma^2(t)}. \tag{15}$$

The latter equality follows from Eq. (12). With this relation we see that Fick's (first) law holds:

$$\frac{D}{P}\frac{\partial P}{\partial r} = -D\frac{2r}{\sigma^2(t)} = -v_r, \tag{16}$$

which confirms our interpretation of $D$ as the diffusion coefficient. It is important to note here that the velocity in Eq. (16) is the velocity of the probability flow. For a large number of particles $P$ is proportional to the concentration, and $v_r$ to the mean velocity at $r \leq r_{max}$. At the end of the previous section we saw that in the case of a single (Brownian) particle moving at $r = r_B$ we find again Eq. (12) with $\sigma(t)$ replaced by $r_B$, so the velocity is $v_{r_B} = 2D/r_B$.

Boltzmann's expression for the thermodynamic entropy, $S = k_B \log \Omega$, provides another connection between the considerations that lead to Eq. (10) and physics. So how is the probability $P$ related to $\Omega$, the number of available microstates per particle? From the derivation of Eq. (10) we conclude that $P$ is proportional to the number of ways a particle that is released from the origin can travel to $\boldsymbol{r}$. In the next section we will use this relation to identify $\Omega$ with $P(\boldsymbol{r})$ in the case of the freely jointed chain attached to the origin. Note however that for a number of freely moving particles released from the origin $P(\boldsymbol{r})$ is proportional to the concentration or density $\rho$. This means that $1/P(\boldsymbol{r})$ can be identified with the available volume of a particle at $\boldsymbol{r}$. In that case we should adopt $\Omega \propto 1/P(\boldsymbol{r})$. This is in agreement with the Sackur-Tetrode equation for the entropy per particle: $S = -k_B \log \rho$, which can be derived from the principle of minimum information.[11] Using Eq. (4), the radial entropic force is then given by

$$F_r = T \frac{\partial S}{\partial r} = -k_B T \frac{\partial \log P}{\partial r} = 2k_B T \frac{r}{\sigma^2} = \frac{v_r k_B T}{D}. \qquad (17)$$

Note that this force is the same as the osmotic force for an ensemble of particles with a spherically symmetrical mass distribution:

$$F_{osm} = -k_B T \frac{\nabla P}{P}. \qquad (18)$$

The osmotic force was defined by Einstein[9] as the force that is in equilibrium with the friction force felt by the Brownian particle. The distribution (10) is not an equilibrium distribution. Note however that by applying an external force equal to $F_{ext} = -F_{osm}$ we can make the distribution (10) stationary.[14,15] Assuming that this force derives from a potential $F_{ext} = -\nabla U(r)$ we see that Eq. (18) yields the Boltzmann distribution,

$$P_{eq}(r) \propto e^{-U(r)/k_B T}. \qquad (19)$$

With $F_{ext} = -F_r$ we find that the potential is the harmonic potential well $U(r) = k_B T r^2/\sigma^2$ and the probability distribution is the Gaussian distribution given in Eq. (10), but now $\sigma$ is fixed.

In deriving $P$ we have used some pieces of information such as the time since the particle was released from the origin. Obviously this information has no effect on the motion of a Brownian particle. So, for a Brownian particle moving at $r$ we will discard this information. We only know that the available volume increases as $r^2$ and we should adopt $\Omega \propto r^2$. This yields Neumann's radial force,

$$F_r = T \frac{\partial S}{\partial r} = \frac{2k_B T}{r}. \qquad (20)$$

This is in agreement with Eq. (17) in view of our conclusion at the end Section II that for a single particle we should adopt $r = \sigma$. Note that Eqs. (17) and (20) both yield a radial force equal to $v_r k_B T/D$.

The entropic force on a Brownian particle with radius $R$ moving at $r$ in a viscous fluid with viscosity $\eta$ is balanced by the drag force given by Stokes' law, $F_{drag}(r) = 6\pi \eta R\, v_r$.

Combining this with Eq. (17) or (20) we find the right expression for the diffusion coefficient:

$$D = \frac{k_B T}{6\pi\eta R}. \tag{21}$$

For particles at $r = \sigma(t)$ we find $\sigma(t)v_{r=\sigma(t)} = 2D$, yielding

$$\langle x^2 \rangle = \tfrac{1}{2}\sigma^2(t) = 2Dt, \tag{22}$$

which is the classical result first obtained by Einstein.[9]

We are used to looking at diffusion as the (mean) movement of particles from a higher to a lower concentration where the drift velocity is directed along the gradient in the density distribution. This description is mathematically correct, but it does not do justice to the fact that the motion of individual particles is not influenced by the properties of the density distribution and it does not describe the motion of an individual particle. On the other hand, the description based on the entropic force, based on minimum information, explains the drift velocity of a single particle as well as the evolution of a collection of particles having a density gradient. This is due to the fact that the probability distribution of each individual particle obeys the diffusion equation, which is a linear differential equation for which the superposition principle holds. The evolution of the ensemble can be described as the evolution of the individual particles combined. If we know all the positions of the particles at time $t_1$ and let the probability distributions for each particle evolve from $t_1$ to $t_2$, we can calculate the probability distribution, and thus the density distribution of the ensemble at $t_2$, by integrating over the probability distributions of the particles. The entropic force at a particular position in the density distribution of an ensemble of particles will be equal to the (osmotic) pressure force given by the gradient in the density distribution of the particles. So the description of diffusion as the result of a radial entropic force on the individual particles appears more general, because it explains the evolution of individual particles as well as an ensemble of particles and avoids the suggestion that the motion of the individual particles in an ensemble is somehow influenced by the other particles.

## IV. THE FREELY JOINTED CHAIN AND HOOKE'S LAW

The freely jointed chain is a very simple model of a polymer consisting of $N_{ch}$ rigid monomers of length $l_{ch}$ whose orientations are independent of each other. The chain forms a pattern similar to that of a random walk and most results from Sections II and III can be applied directly to find the entropic force on such a chain. When one end of the chain is fixed to the origin, the probability distribution for the other endpoint $\boldsymbol{r}$ of the chain is given by Eq. (10), with $\sigma^2 = \tfrac{2}{3}N_{ch}l_{ch}^2$. The difference with the radial force on the Brownian particle is in the definition of $\Omega$. Remember that $P(\boldsymbol{r})$ could be identified with the number of ways a particle released from the origin can arrive at $\boldsymbol{r}$. This number is equal to the number of configurations of a chain with one end fixed at the origin and the other end at $\boldsymbol{r}$. So now the resulting force is given by Eq. (17) and has sign opposite to that of the radial force in Eq. (20). We find, for a chain embedded in a medium with temperature $T$,

$$F^{ch} = T\frac{\partial S}{\partial r} = k_B T \frac{\partial \log P}{\partial r} = -k_B T \frac{3r}{N_{ch}l_{ch}^2} = -ar, \tag{23}$$

consistent with Hooke's law. The chain behaves as a spring, or harmonic oscillator, with force proportional to its length. It is interesting to see what happens if we attach a Brownian particle to the end of this spring. The balance of the two opposite radial forces in Eqs. (23) and (20) yields an equilibrium distance: $\sigma^2 = 2\langle x^2 \rangle = 2k_BT/a$. This is in agreement with the classical result for the position of a Brownian particle in a harmonic potential well. Note that, as explained below Eq. (11), the radial entropic force exerted by the Brownian particle is due to *tangential* motions of the particle, which increase the distance from the origin.

Equation (23) is valid for $r \ll N_{ch}l_{ch}$, or $F^{ch} \ll 3k_BT/l_{ch}$. The relative elongation of a chain that is pulled by an external force $F^{ext}$ equal to $-F^{ch}$ is given by $r/L_{ch} = r/N_{ch}l_{ch} = F^{ext}l_{ch}/(3k_BT)$. This must be $\leq 1$, so what happens at large values of $F^{ext}l_{ch}/(k_BT)$? In the next section we will see that the picture is incomplete because we have neglected another entropic force. Incorporation of this force will give us a more general expression, which is also valid for high values of the relative elongation.

## V. THE TANGENTIAL ENTROPIC FORCE

A tangential force becomes apparent as soon as we choose the angular coordinates of a Brownian particle moving at position $\boldsymbol{r} \neq 0$. Let us choose a particular direction $\boldsymbol{e_D}$ ("North Pole") defined by $\theta = 0$ (choosing the *z*-axis along $\boldsymbol{e_D}$; see Fig. 1). This could be the direction of an external force, or—in the case of the drunkard's walk—the direction home. Then it is likely that, starting from the center of the sphere, the drunkard will appear somewhere around the equator rather than at his house at the North Pole simply because the circles of latitude become larger towards the equator. This "force" becomes apparent also when a magnetic needle in a magnetic field is placed in a hot bath. A temperature-dependent entropic force tends to drive the needle away from the direction of the magnetic field. This phenomenon provides a way to construct a magnetic thermometer.

Let us find an expression for this tangential entropic force in the same way as we did for the radial force. Suppose that the particle is fixed at a distance $r$ from the origin and is only allowed to move over the sphere with this radius. The work done by an external force $\boldsymbol{F}$ (for example a magnetic field) working in the direction $\theta = 0$ to bring the particle (or magnetic needle) from $\theta_1$ to $\theta_2$ is $\Delta U = \int_{\theta_1}^{\theta_2} F\, r\, \sin\theta\, d\theta = F\, r\, \Delta \cos\theta$. This is the work done by a torque $M = |\boldsymbol{M}| = |\boldsymbol{F} \times \boldsymbol{r}| = F_{tang} r$. We should introduce this work $U(\theta) = F\, r(1 - \cos\theta)$ in the microcanonical ensemble. The equilibrium of forces, or torques, demands that

$$\frac{d}{d\theta} S(E + F\, r(1 - \cos\theta), \theta) = 0, \qquad (24)$$

yielding

$$T \frac{d}{d\theta} S = F\, r\, \sin\theta = |\boldsymbol{F} \times \boldsymbol{r}| = M. \qquad (25)$$

Next we derive the probability density for the latitude distribution consistent with our lack of information about the latitude $\theta$, where $\theta$ is the angle between $\boldsymbol{r}$ and $\boldsymbol{e_D}$. We follow the same reasoning as in the derivation of Eq. (8). Our *a priori* distribution $q(\theta)$ should be proportional to the circumference of the latitude circle $2\pi\, r\sin\theta$, so $q(\theta)d\theta = \frac{1}{2}d\cos\theta$. The probability of the new distribution $p(\theta)d\cos\theta$ is then given by Eq. (6). We adopt the Lagrange multiplier $\tau$ to incorporate the constraint that the mean value $\langle \cos \theta \rangle$ is finite and find that $\log \text{Prob}[p(\theta)]$ has a maximum for the normalized distribution:

$$p(\theta) d\cos\theta = \frac{\tau\, e^{\tau \cdot \cos\theta}}{e^\tau - e^{-\tau}} d\cos\theta. \qquad (26)$$

The mean value $\langle \cos \theta \rangle$ for this distribution is related to $\tau$ by

$$\langle \cos\theta \rangle = \frac{e^\tau + e^{-\tau}}{e^\tau - e^{-\tau}} - \frac{1}{\tau} = \coth(\tau) - \frac{1}{\tau} = \Lambda(\tau), \qquad (27)$$

Where $\Lambda(\tau)$ is known as the Langevin function, with the properties $\Lambda(\tau) \cong \tau/3$ for small $\tau$ and $\Lambda(\tau) \to 1$ for large $\tau$.

The entropy can be defined as $S = k_B \log p(\theta)$, and so the strength of the entropic torque at radius $r$ is given by

$$M = T\frac{dS}{d\theta} = k_B T \frac{1}{p(\theta)} \frac{dp(\theta)}{d\theta} = k_B T \tau \sin\theta. \qquad (28)$$

Combining Eqs. (25) and (28), we have equilibrium for $F = k_B T \tau / r$, which is $\cong 3 k_B T \langle \cos\theta \rangle / r$ for small $\tau$.

First let us apply an external force in the case of the free chain. Define $\theta = 0$ as the mean direction of the chain, so $\langle \theta \rangle = 0$ for the monomers. The angle distribution of the monomers is given by Eq. (26), so the relative elongation is $r/L_{ch} = r/N_{ch} l_{ch} = \langle \cos\theta \rangle = \Lambda(\tau)$. For high values of $\tau$ the relative elongation goes to 1 as it should. The applied external force $\boldsymbol{F}^{ext}$ pulls in the direction $\theta = 0$. The tangential component of this force on an individual monomer $i$ of length $l_{ch}$ and orientation $\theta_i$ is $F^{ext} \sin\theta_i$. For equilibrium this should be equal to $F_{tang} = k_B T \tau \sin\theta_i / l_{ch}$, implying $\tau = F^{ext} l_{ch} / k_B T$. For low values of $\tau$ the relative elongation $r/L_{ch} \cong \Lambda(\tau)$ goes to $\tau/3 = F^{ext} l_{ch} / 3 k_B T$, in agreement with Eq. (23).

Next we consider the entropic force on electric and magnetic dipoles in a thermal bath. The force $\mu B \sin\theta$ on a magnetic dipole $\boldsymbol{\mu}$ making an angle $\theta$ with a magnetic field will be in equilibrium with this entropic force when $\tau = \mu B / k_B T$. For a large number of magnetic dipoles in a magnetic field the mean projection of the magnetic moments along the direction of the magnetic field, or the induced dipole moment, is given by

$$\langle \boldsymbol{\mu} \rangle = \mu \langle \cos\theta \rangle = \frac{\mu^2 B}{k_B T} \Lambda(\tau) \approx \frac{\mu^2 B}{3 k_B T} \quad \text{for small } \tau. \qquad (29)$$

This result is known as the Langevin-Debye equation in the case of electric dipoles in an electric field, and Curie's law in the case of magnetic dipoles in a magnetic field. The magnetic thermometer mentioned above is based on Curie's law, which can be interpreted as a direct result of the tangential entropic force.

## VI. SUMMARY AND DISCUSSION

The main purpose of this paper is to show how the concept of entropic forces can be illustrated in Brownian motion (or diffusion). The results obtained are not new, but they are derived in a somewhat different way, which may help to increase our understanding of the concept of entropic forces in the case of Brownian motion. In a spherically symmetric system the Brownian particles feel a radial entropic force. The probability distribution of a Brownian particle that is released from the origin of a coordinate system is derived from the principle of minimum information. The probability distribution as a function of the distance to the origin consists of two factors: 1) an exponential factor for the total number of paths between the origin and the position of the particle; and 2) a factor for the increase in the number of volume elements with radius. The osmotic force for a collection of particles with a given concentration gradient follows from the first factor. The radial entropic force experienced by a single Brownian particle at distance $r$ from the origin follows from the second factor. These forces are equal at $r = \sigma(t)$. This approach yields the Einstein relation and the diffusion equation. Diffusion can be regarded as a phenomenon that is driven, or caused, by an entropic force on the individual particles in the ensemble.

The path of a Brownian particle (with constant step length) is similar to the configuration of a chain of freely moving monomers. It is then straightforward to calculate the entropic force on the chain, which is consistent with Hooke's law. For a complete description, however, we have to take the transverse component of the entropic force (or torque) into account. This force arises as soon as we choose a particular direction ($\theta = 0$), for instance the direction of an external force. It is a tangential entropic force (or torque) working in the direction of the equator ($\theta = \pi/2$). The derivation of the probability distribution and the tangential entropic force are again based on the principle of minimum information. The results are applied to Hooke's law and to magnetic (or electric) dipoles in a magnetic (or electric) field and shown to reproduce Curie's law (and the Langevin-Debye law).

It may seem odd to derive a physical force from the minimum information principle, which states that the distribution should be consistent with our lack of knowledge of the positions of the particles. The source of confusion lies in the close connection between the concepts of entropy in thermodynamics and in information theory. The force results from the microscopic fluctuations in a thermodynamic system, which are driving the system towards states with larger probability, or higher entropy. Increase of entropy goes along with decrease of information: as the system evolves, the number of bits required to describe a particular state increases and the information we have about an initial state of the system becomes less adequate to describe the system. On a macroscopic scale this increase in entropy can be described as an (emerging) entropic force using Boltzmann's law. Such forces are just as real as the physical forces they counteract, such as an external pulling force in Hooke's law or a magnetic field in the case of Curie's law. We have to apply an external force to prevent Brownian particles from drifting away from their original position or from a particular direction. Brownian motion thus provides a basic example of entropic forces emerging from microscopic fluctuations. It is amazing that such forces may ultimately explain Newton's second law, gravity, and perhaps even the dark energy that is pulling the universe apart.

**Acknowledgments**

I thank E. Verlinde for his support, H. van Beijeren and especially Richard M. Neumann for comments and discussions, and Bert van der Mey for the illustration (Fig. 1). I thank the Leiden Observatory for its hospitality.

* nico.roos@hccnet.nl

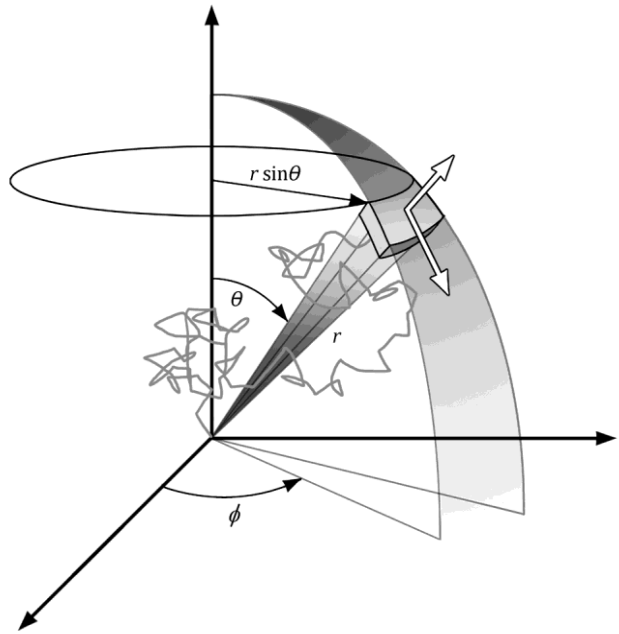

Figure 1. The entropic forces (white arrows) on a Brownian particle in a volume element for a spherically symmetric system.